\documentclass[%
 aps,
 amsmath,amssymb,
 reprint,
]{revtex4-1}

\usepackage{graphicx}
\usepackage{dcolumn}
\usepackage{bm}
\usepackage{hyperref} 
\usepackage{verbatim} 
\usepackage[utf8]{inputenc}
\usepackage[T1]{fontenc}
\usepackage{mathptmx}

\usepackage{xcolor}

\preprint{APS/123-QED}
\begin{document}

\title{Phase-tunable thermoelectricity in a Josephson junction} 

\author{G. Marchegiani}%
 \email{giampiero.marchegiani@nano.cnr.it}
\affiliation{
NEST Istituto Nanoscienze-CNR and Scuola Normale Superiore, I-56127 Pisa, Italy}%
\author{A. Braggio}%
 \email{alessandro.braggio@nano.cnr.it}
\affiliation{
NEST Istituto Nanoscienze-CNR and Scuola Normale Superiore, I-56127 Pisa, Italy}%
\author{F. Giazotto}%
 \email{francesco.giazotto@sns.it}
\affiliation{
NEST Istituto Nanoscienze-CNR and Scuola Normale Superiore, I-56127 Pisa, Italy}

\date{\today}

\begin{abstract}
Superconducting tunnel junctions constitute the units of superconducting quantum circuits and are massively used both for quantum sensing and quantum computation. In previous works, we predicted the existence of a nonlinear thermoelectric effect in a electron-hole symmetric system, namely, a thermally biased tunnel junction between two different superconductors, where the Josephson effect is suppressed. In this paper we investigate the impact of the phase-coherent contributions on the thermoelectric effect, by tuning the size of the Josephson coupling changing the flux of a direct-current Superconducting Quantum Interference Device (dc-SQUID). For a suppressed Josephson coupling, the system generates a finite average thermoelectric signal, combined to an oscillation due to the standard ac Josephson phenomenology. At large Josephson couplings, the thermoelectricity induces an oscillatory behaviour with zero average value of the current/voltage with an amplitude and a frequency associated to the Josephson coupling strength, and ultimately tuned by the dc-SQUID magnetic flux. In conclusion, we demonstrate to be able to control the dynamics of the spontaneous breaking of the electron-hole symmetry. Furthermore, we compute how the flux applied to the dc-SQUID and the lumped elements of the circuit determine the frequency of the thermoelectric signal across the structure, and we envision a frequency modulation application.
\end{abstract}


\maketitle 

\section{Introduction}
\label{Introduction}
The investigation of thermal transport in micro/nanoscale systems has attracted a growing interest in the last few decades~\cite{Goodson2006,GiazottoRMP2006,Muhonen2012,DiVentraRMP2011,CahillReviewII2014,BergfieldMolecular,WangEPJB2008,Pop2010}, and is expected to have an impact on the performance of modern quantum technologies~\cite{Riedel_2017,Acn2018}. Heat dissipation is a key factor and limits the performance of classical computation platforms, but it is even more crucial in multi-qubit technology, where low operating temperatures further limit the heat exchange. Hybrid superconducting junctions~\cite{Tinkham2004,barone1982physics} are ideal platforms for quantum devices~\cite{Kurizki3866,PekolaReview2015,Wendin2017,Krantz2019}, due to the well-established fabrication techniques and a precise modeling of the coherent electronic transport. In particular, they offer a tight control over thermal currents, with applications to electronic solid state cooling~\cite{GiazottoRMP2006,Muhonen2012}, phase-coherent modulation of thermal currents~\cite{FornieriReview,Hwang2020}, and quantum sensing~\cite{PekolaRMP85}. In the last few years, they have been also been extensively investigated for thermoelectricity, when superconductors are used in combination with ferromagnetic elements~\cite{BelzigTEPRL,Ozaeta2014,Kolenda2017,KolendaPRL,VirtanenReviewTE}, and the interplay between thermoelectricity and the superconducting phase is being established~\cite{ThermophasePRL,GlazmanPRB99,KampSothmannPRB,KalenkovPRB2020,BlasiPRL}. In Refs.~\onlinecite{MarchegianiPRL,MarchegianiPRB}, we predicted an unexpected nonlinear thermoelectric effect occurring in a system with electron-hole (EH) symmetry, paradigmatically a tunnel junction between two different Bardeen-Cooper-Schrieffer (BCS) superconductors. We observed that thermoelectricity arises due to a spontaneous breaking of EH symmetry determined by the electrode with the larger gap to have the higher temperature~\cite{MarchegianiPRL,MarchegianiPRB}. In the discussion, we focused only on the quasiparticle transport across the junction and we assumed to be able to suppress completely any phase-dependent contribution associated to the Josephson effect~\cite{barone1982physics}. Purpose of this work is to investigate in details the impact of the phase-dependent terms on the thermoelectric behavior. As we will show below, the generation of a finite thermoelectric voltage is still possible in the presence of Josephson contributions, but an additional oscillating behaviour is generated in accordance to the Josephson effect. We demonstrate that the spontaneous generation of a thermoelectric voltage/current can be controlled by tuning the strength of the phase dependent terms, which, for the setup we consider, can be modulated by changing the magnetic flux in a Superconducting Quantum Interference Device (SQUID)~\cite{barone1982physics,clarke2006squid}. Moreover, we discuss the impact of the Josephson terms on the whole dynamics of the system. In particular, the frequency and the amplitude of the thermoelectric-induced oscillations are numerically computed, and approximate expressions are obtained in some limiting cases.

\section{Model and Results}
\subsection{Charge transport and thermoelectricity in a superconducting junction}
\begin{figure}[tp]
	\begin{centering}
		\includegraphics[width=\columnwidth]{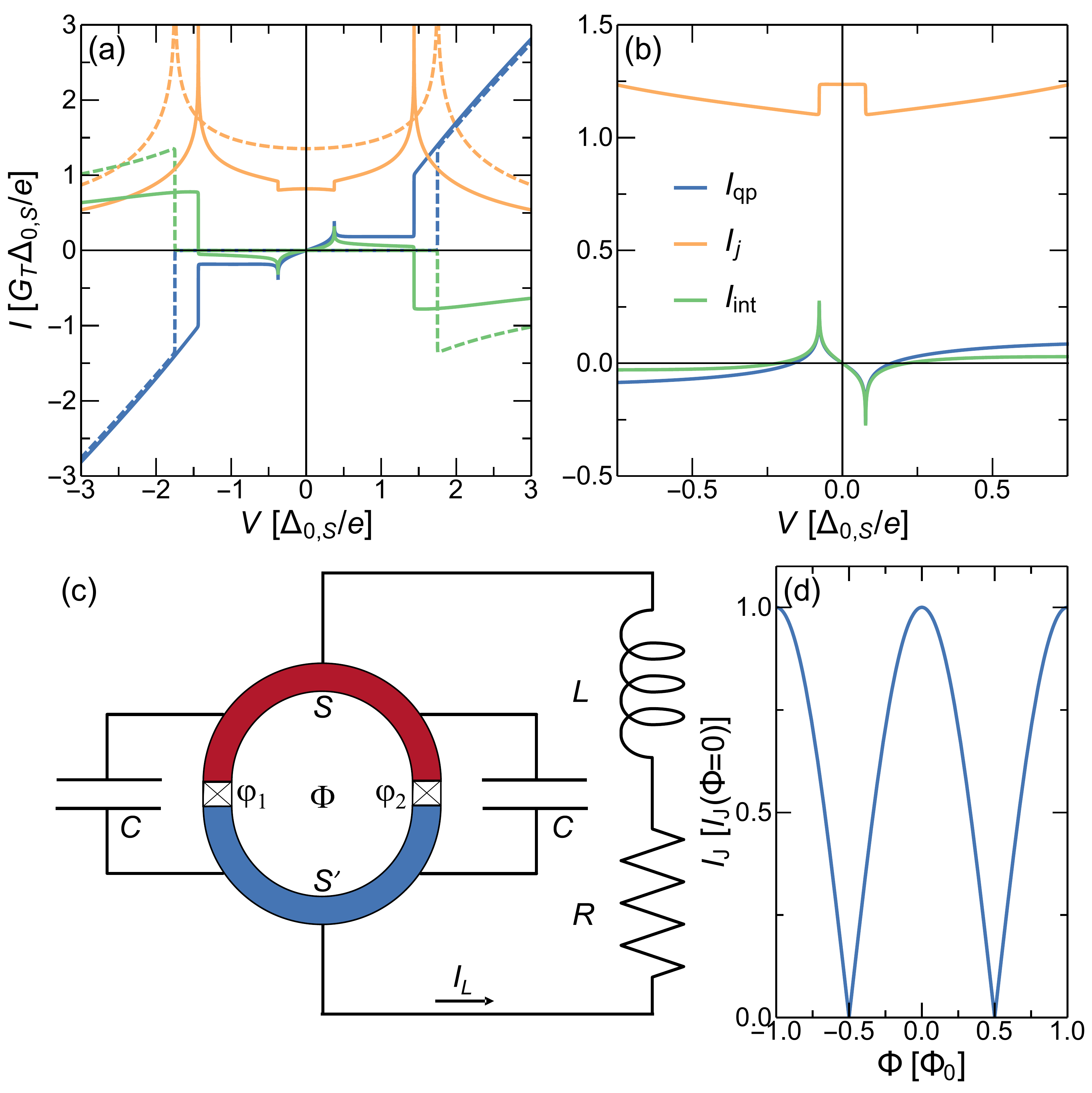}
		\caption{\textbf{(a)} Voltage evolution of the three contributions $I_{\rm qp},I_{\rm int}, I_j$ to the current in a tunnel junction between two BCS superconductors (SIS' junction) for $T_{S}=T_{S'}=T$, and $r=0.75$. Dashed curves give the low temperature behaviour $T\to 0$, whereas the solid curves are displayed for $T=0.6T_{c,S}$. \textbf{(b)} Subgap voltage evolution of $I_{\rm qp},I_{\rm int}, I_j$ in the presence of a thermal gradient $T_{S}=0.7 T_{ c,S}$, $T_{S'}=0.01 T_{ c,S}$. The absolute negative conductance in the quasiparticle and the interference contributions denotes a thermoelectric behaviour. \textbf{(c)} Circuit scheme of the proposed system. A superconducting ring made of two different superconductors, interrupted by two Josephson junctions, is connected in series with an external RL circuit. A thermal bias is imposed between the two superconductors. \textbf{(d)} Flux evolution of the Josephson current for two symmetric junctions.
		}
		\label{Fig1}
	\end{centering}
\end{figure}

The charge current in a tunnel junction between two superconductors depends both on the phase bias ($\varphi$) and the voltage ($V$) applied to the junction, as first predicted by Josephson~\cite{JOSEPHSON1962251,barone1982physics,Harris1974}
\begin{equation}
I(V,\varphi)=I_{\rm qp}(V)+I_{j}(V)\sin\varphi+I_{\rm int}(V)\cos\varphi,
\label{eq:IVphi_characteristic}
\end{equation}
where $I_{\rm qp}$ is the quasiparticle contribution, $I_{j}$ is associated with Cooper pairs tunneling, and $I_{\rm int}$ gives the interference contribution associated with breaking and recombination process of Cooper pairs on different electrodes of the junction~\cite{GuttmanPRB55}. The explicit expressions of $I_{\rm qp},I_{j},I_{\rm int}$ are given in Appendix A, being well-known in the literature~\cite{Harris1974,barone1982physics,GulevichPRB96}.
The current obeys the symmetry $I(V,\varphi)=-I(-V,-\varphi)$. Thus, $I_{\rm qp},I_{\rm int}$ are odd in $V$ and represent the dissipative (or active in the presence of thermoelectricity) components of the current, whereas the function $I_{\rm j}(V)$ is even and corresponds to a purely reactive contribution~\cite{HarrisPRB11,Zwerger1983}. Indeed, in the presence of a phase bias, the junction can support an equilibrium (non-dissipative) current even for $V=0$, $I=I_{c}\sin\varphi$ (dc Josephson effect), where $I_c=I_j(V=0)$ is called critical current. 
At a finite voltage $V\neq 0$, the phase across the junction oscillates in time according to the Josephson equation (AC Josephson effect)
\begin{equation}
\frac{d\varphi}{dt}=\frac{2eV}{\hbar}
\label{eq:ACJosephson}
\end{equation}
where $\hbar$ is the reduced Planck constant, and $-e$ is the electron charge. Namely, for a constant bias $V(t)=V_0$ the phase dependent terms oscillates in time with Josephson frequency $f_j=|V_0|/\Phi_0$ and zero average value, where $\Phi_0=\pi\hbar/e\sim 2$ fWb is the flux quantum. In this case, the dc response of the junction is given by the quasiparticle contribution only. For our purposes, we consider a junction between two different superconductors ($S,S'$), and introduce an asymmetry parameter $r=\Delta_{0,S'}/\Delta_{0,S}=T_{c,S'}/T_{c,S}$, where $\Delta_{0,i}=1.764 k_B T_{c,i}$ (with $i={S,S'}$) is the zero-temperature order parameter, and $T_{\rm c,i}$ is the critical temperature of the $i$ electrode.
Figure~\ref{Fig1}a displays the voltage dependence of the three contributions when the electrodes have equal temperatures $T_S=T_{S'}=T$ and $r=0.75$. In the low temperature limit $T\ll T_{c,S}$ (dashed lines), both $I_{\rm qp}$ and $I_{\rm int}$ are strongly suppressed for $|V|<(\Delta_{ 0,S}+\Delta_{0,S'})/e$ and finite at higher voltage. Note that $I_{\rm \rm qp}$ is positive and monotonically increasing for $V>(\Delta_{0,S}+\Delta_{0,S'})/e$, where it asymptotically reads $I_{\rm qp}=G_T V$ ($G_T$ is the normal state conductance). 
On the other hand, $I_{\rm int}$ is negative and monotonically decreasing for $V>(\Delta_{0,S}+\Delta_{0,S'})/e$. The Cooper pairs term  $I_{\rm j}$ has a more complex evolution, and it is monotonically increasing for $0<V<(\Delta_{0,S}+\Delta_{0,S'})/e$, where it diverges (the divergence is smoothed with the introduction of a finite phenomenological parameter $\Gamma$, see Appendix A), and it is monotonically decreasing at higher voltages. In the same plot, we display the evolution also for a finite value of the temperature, i.e., $T=0.6 T_{c,S}$. While, the overall behaviour of the curves is similar, $I_{\rm qp}$, $I_{\rm int}$ are now finite and displays a positive conductance $G_i=I_{i}/V>0$, with $i=\{\rm qp,int\}$ at subgap voltages $|V|<[\Delta_{S}(T)+\Delta_{S'}(T)]/e$, showing a nonlinear evolution characterized by a peak for $V=\pm V_p=\pm|\Delta_{S}(T)-\Delta_{S'}(T)|/e$, due to the matching of the BCS singularities in the density of states of the two superconductors~\cite{barone1982physics}. 
With increased temperature, the Cooper pairs term is reduced due to the monotonically decreasing evolution of the superconducting gaps [$\Delta_i(T)\leq \Delta_{0,i}$ ]. For the same reason, the voltage value where the Cooper pairs term has the peak and the other contributions have a sharp jump, i.e., $V=[\Delta_{S}(T)+\Delta_{S'}(T)]/e$ is reduced with respect to the low temperature limit.  

Since we are interested in the description of thermoelectric phenomena, we consider a situation where a temperature difference is established between the electrodes, namely $T_{S}\neq T_{S'}$. Note that in the absence of a temperature bias, i.e., $T_{S}= T_{S'}$, the behaviour of the junction is purely dissipative, since $(I_{\rm qp}+I_{\rm int}\cos\varphi)V>0$ for every $\varphi$~\cite{Zwerger1983}, as required by the second law of thermodynamics~\cite{MarchegianiPRL}. Conversely, with a thermal gradient it is possible to have a thermoelectric power generation with positive entropy production. In particular, a thermoelectric behaviour is characterized by a positive thermoelectric power $\dot W=-IV>0$ produced by the junction. As discussed above, this definition mainly applies to the even-$\varphi$ component of the current, i.e., $I_{\rm qp}+I_{\rm int}\cos\varphi$, since the Cooper pairs term $I_{j}\sin\varphi$ essentially describes a reactive component. In Ref.~\onlinecite{MarchegianiPRL}, we predicted the existence of a thermoelectric effect at subgap voltages in the quasiparticle component, namely we demonstrated that we can have $I_{\rm qp}V<0$ for small values of $V$, provided that the superconductor with the larger gap is heated up [in our notation $T_{S}>T_{S'}$ and $\Delta_{S}(T_{S})>\Delta_{S'}(T_{S'})$, since we assumed $\Delta_{0,S}>\Delta_{0,S'}$]. 
This is shown in Fig.~\ref{Fig1}b, where the subgap evolution of $I_{\rm qp},I_{\rm int},I_j$ is displayed for $T_S=0.7 T_{c,S}$ and $T_{S'}=0.01 T_{c,S}$. In particular, at a low voltage the quasiparticle curve displays a peculiar negative conductance $G_{\rm qp}(V)=I_{\rm qp}(V)/V<0$ and hence finite thermoelectric power $\dot W=-I_{\rm qp}V>0$. In the absence of phase-dependent terms, the negative value of $G_{\rm qp}$ for $V\to 0$ implies a spontaneous breaking of electron-hole symmetry. This leads to the generation of a thermoelectric voltage due to the existence of finite voltage values $\pm V_S$ where the current is zero [$I_{\rm qp}(V_S)=0$], as discussed in Refs.~\onlinecite{MarchegianiPRL,MarchegianiPRB}. 

Consider now the phase dependent terms. Interestingly, the interference term (green) behaves similarly to the quasiparticle term, also showing a negative conductance $G_{\rm int}(V)=I_{\rm int}(V)/V<0$ around the origin~\footnote{This implies that $I_{\rm int}$ can provide thermoelectric power during the time evolution, as discussed in the following sections.}. In particular, the zero-bias value of the differential conductance reads $G_{\rm 0,int}=G_{\rm 0,qp}\Delta_{S}(T_S)/\Delta_{0,S'}$ in the limit $T_{S'}\to 0$ (see Appendix A). The Cooper pairs term (yellow) is roughly constant for $V<V_p=[\Delta_S-\Delta_{S'}]/e$, where it sharply decreases, and rise monotonically at higher voltages. Similar jumps are observed also in the temperature evolution of the critical current~\cite{GuarcelloPRApplied}. Note that the size of the Cooper pairs term, which is finite at $V=0$, is quite larger than the quasiparticle contribution. As a consequence, it is reasonable to expect the Josephson current to have a  potentially crucial impact on the features of the thermoelectric effect.
\subsection{Circuit dynamics modeling}
In order to describe the impact of the phase-coherent contributions on the thermoelectricity of the junction, we consider a system where the size of $I_j$, $I_{\rm int}$ can be externally tuned. More precisely, we investigate the circuit displayed in Fig.~\ref{Fig1}c. The system features a superconducting ring made of two different superconductors, which are coupled by two tunnel junctions. This configuration is known in the literature as the direct-current Superconducting Quantum Interference Device (dc-SQUID)~\cite{barone1982physics,clarke2006squid}. We assume that the superconducting ring is connected to an external circuit, which constitutes an idealized model for the electrical environment described in terms of lumped elements, i.e., an inductance $L$ and a load $R$. Each of the two junctions displays the nonlinear current-voltage characteristic $I(V,\varphi)$ of Eq.~\eqref{eq:IVphi_characteristic}, so that the total current in the dc-SQUID reads
\begin{equation}
I_{\rm SQ}=I_1(V,\varphi_1)+ I_2(V,\varphi_2)=I(V,\varphi_1)+\alpha I(V,\varphi_2)
\label{eq:squidnaive}
\end{equation}
where $\varphi_{1}$ and $\varphi_{2}$ are the phase-differences across the two junctions, and $\alpha=G_{T2}/G_{T1}$ is the ratio between the conductances of the two junctions in the normal state. For simplicity, in the theoretical modeling we will consider a fully symmetric SQUID ($\alpha=1$), even if the results can be properly extended to asymmetric junctions~\footnote{In the numerical calculations presented in this work we include a small asymmetry, namely $G_{T1}/G_{T2}=\alpha=0.95$, to give a more realistic picture.}. For a proper description of the dynamics, we need to consider also the capacitance $C$ of each of the two junctions.
Due to the ring geometry, the superconducting phase differences across the two junctions are related by the fluxoid quantization, namely,
\begin{equation}
\varphi_1(t)-\varphi_2(t)+2\pi\Phi/\Phi_0=2\pi n,\quad n\in \mathbb{Z}
\label{eq:fluxquantization}
\end{equation}
where $\Phi$ is the total flux out of plane of the superconducting ring. The magnetic flux $\Phi$ coincides with the flux applied externally, since we assume the self-inductance of the ring negligible. By minimizing the free energy of the SQUID with respect to the superconducting phases $\varphi_1,\varphi_2$, and using the constraint of the flux quantization Eq.~\eqref{eq:fluxquantization}, we can rewrite the expression of Eq.~\eqref{eq:squidnaive} as
\begin{equation}
I_{\rm SQ}=2I_{\rm qp}(V)+
2\left|
\cos\left(\pi\frac{\Phi}{\Phi_0}\right)\right|[I_{j}(V)\sin\tilde\varphi+
I_{\rm int}(V)\cos\tilde\varphi],
\label{eq:Isquid}
\end{equation}
where $\tilde\varphi=(\varphi_1+\varphi_2)/2$.
The circuit dynamics is finally expressed by~\footnote{We stress that Eq.~\eqref{eq:IVphi_characteristic}, which leads to Eq.~\eqref{eq:Isquid}, holds exactly only for a constant voltage bias $V$. However, the expression still gives a good approximation for a time-dependent voltage bias $V(t)$ in the adiabatic regime, valid for our parameter values, as discussed in Appendix A.} 
\begin{equation}
\begin{cases}
I_{\rm L}=2C \dot V+I_{\rm SQ}(V,\tilde\varphi,\Phi) \\
V=-L \dot I_{\rm L}-RI_{\rm L}\\
\dot {\tilde\varphi}=2eV/\hbar,
\end{cases}
\label{sys:circuit}
\end{equation}
which is an autonomous non-linear system of differential equations in the three variables $\tilde\varphi,V,I_{\rm L}$. The first equation in Eq.~\eqref{sys:circuit} gives the current conservation in the circuit: the current $I_L$ which flows in the inductor $L$ and in the load $R$ is the sum of the current in the capacitances (first term in the right side) and in the two junctions (second term in the right side). The second equation in Eq.~\eqref{sys:circuit} gives the Kirchhoff voltage rule in the circuit: the voltage $V$ across the SQUID is equal to the sum of the voltage drops in the inductance and in the load. The last identity in Eq.~\eqref{sys:circuit} follows from the Josephson relation between the phase bias and the voltage bias in a Josephson junction of Eq.~\eqref{eq:ACJosephson}. 
As can be seen from Eq.~\eqref{eq:Isquid}, the absolute strength of the phase-dependent contributions can be fully tuned by varying the magnetic flux $\Phi$, as shown in Fig.~\ref{Fig1}d for the Cooper pairs term $I_j$ at $V=0$. In particular, the evolution of the phase-coherent contributions is periodic with period $\Phi_0$: the Josephson current is maximum for $\Phi=n\Phi_0$ (with $n\in\mathbb{Z}$) and it is exactly zero for $\Phi=(1/2+n)\Phi_0$~\footnote{This total destructive interference is due to the perfect symmetry ($\alpha=1$) between the two junctions of the SQUID. However, it is possible to generalize the discussion for asymmetric junctions. For instance, the flux-evolution of the critical current in the SQUID reads $I_c=I_{j,SQ}^{\rm max}(V=0)=I_J(V=0)\sqrt{1+\alpha^2 +2\alpha \cos(2\pi\Phi/\Phi_0)}$}. For this reason, we will consider the evolution only in a single period $\Phi\in[0,\Phi_0]$.

\subsection{Flux modulation of the dc-thermoelectricity} 
We wish to investigate if the thermovoltage can be generated in the presence of the phase-dependent terms. First, we consider situations where these terms are suppressed, which mainly happens for $\Phi\sim(n+1/2)\Phi_0$ (with $n\in \mathbb{Z}$). Indeed, for $\Phi=(n+1/2)\Phi_0$ the phase-coherent contributions are zero. In this case, the dynamics of the variables $V,I_{\rm L}$ is independent on $\tilde\varphi$. This limit corresponds to the one previously discussed in Ref.~\onlinecite{MarchegianiPRL}. In particular, the stationary time-independent solutions are obtained by solving the implicit equation~\footnote{For asymmetric junctions $\alpha\neq 1$, one has to replace $2\to 1+\alpha$.}
\begin{equation}
I_L=2I_{\rm qp}(\bar V)=-\bar V/R.
\label{eq:implicit}
\end{equation} 
For a dissipative junction, $I_{\rm qp}V>0$, and the only solution of Eq.~\eqref{eq:implicit} is $V=0$ (and thus $I_L=0$). However, in the presence of a thermoelectric effect, the behaviour of the system depends on the size of the load $R$~\cite{MarchegianiPRL,MarchegianiPRB}. That is, for $R<V_p/(2I_{p})$, the system may display a oscillatory behaviour with zero average value of $I_{L}$ and $V$. Conversely, for $R>V_p/(2I_{p})$, the system admits stationary and time-independent solutions $(V,I_{\rm L})=(\bar V,2I_{\rm qp}(\bar V))$, where $|\bar V|>V_p$. Note that, due to EH symmetry, each positive solution $\bar V>0$ has a corresponding solution $-\bar V<0$. As a consequence, the system approaches either $\bar V$ or $-\bar V$ in the steady-state, depending on the specific initial condition~\cite{MarchegianiPRL}.

\begin{figure}[tp]
	\begin{centering}
		\includegraphics[width=\columnwidth]{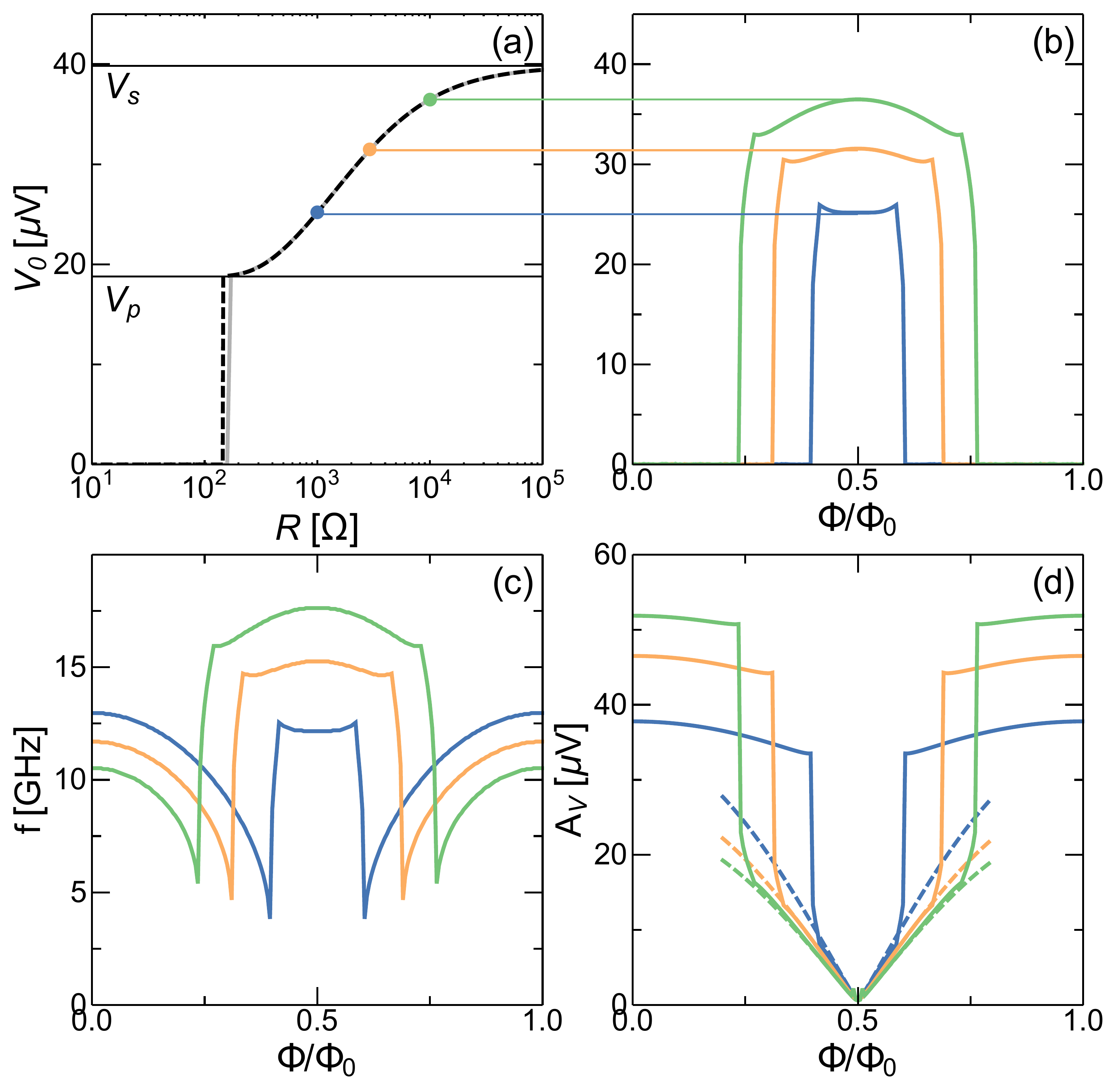}
		\caption{\textbf{(a)} Mean thermoelectric voltage across the SQUID in the absence of the Josephson effect, as a function of the load $R$. The thermo-voltage obtained by numerical solution of the system of Eq.~\eqref{sys:circuit} (solid) is compared to the self-consistent solution of Eq.~\eqref{eq:implicit} (dashed). \textbf{(b)} Flux evolution of the mean thermoelectric voltage for selected values of $R$ (marked by the colored points in the panel a)). \textbf{(c)} Flux dependence of the frequency of the oscillations in the steady-state.  \textbf{(d)} Flux control of the amplitude of the oscillation in the steady-state. Parameters are: $G_{T}^{-1}=1{\rm k}\Omega$, $L=1$ nH,  $C=100$ fF, $T_{S}=0.7T_{c,S}$, $T_{S'}=0.01T_{c,S}$, $T_{c,S}=1.6 K$, and $r=0.75$.}
		\label{Fig2}
	\end{centering}
\end{figure}

 Here, and in the rest of this work, we a set of realistic parameters, for an aluminum-based SQUID with $T_{c,S}=1.6$ K (and thus $\Delta_{0,S}= 1.764 k_B T_{c,S}\sim 240\mu$eV) and $G_T=(1\textrm{k}\Omega)^{-1}$. We consider the thermoelectric situation displayed in Fig.~\ref{Fig1}b for scaled quantities, where $r=0.75$ and $V_p\sim 0.08 \Delta_{0,S}\simeq 19 \mu$eV. Figure~\ref{Fig2}a displays the absolute value of the stationary value of the thermoelectric voltage $V_0=|V|$ (solid) as a function of the load $R$ computed through numerical solution of the system of Eq.~\eqref{sys:circuit} (see the discussion on Sec.~\ref{sec:Fluxevolution}) and the thermoelectric voltage obtained by solving the implicit equation Eq.~\eqref{eq:implicit} (dashed). Note that the two quantities coincide, except for a very narrow range $150\Omega\leq R\leq 200 \Omega$ where the solution of the implicit equation is different from zero, while the result of the numerical integration is zero. This small different behaviour is associated to the stability of the $V=0$ solution of Eq.~\eqref{eq:implicit} and will be discussed in more details after. As discussed above, the thermoelectric voltage is zero for low values of $R$ and it is finite (and larger than $V_p$) and monotonically increasing for $R> V_p/[2I(V_p)]\sim 150 \Omega$. In the limit $R\to\infty$, the thermoelectric voltage approach the Seebeck voltage $V_S$, i.e., the zero-current solution $I(V_{S})=0$ with finite voltage bias $V_S\neq 0$. 
\subsubsection{Small Josephson contribution}
In the presence of a small Josephson current, the picture described above is expected to be slightly modified. Indeed, in the presence of a finite voltage $\bar V$, the phase evolves in time $\tilde\varphi(t)\sim 2e\bar Vt/\hbar$ due to the ac Josephson effect and so an oscillating term $\delta V(t)$ with characteristic frequency $f_{\bar V}=\bar V/\Phi_0$ is superimposed to the dc thermoelectric voltage, i.e., $V(t)\sim\bar V+\delta V(t)$. In order to compute the perturbative contribution $\delta V(t)$, we consider the first equation in Eq.~\eqref{sys:circuit}, $\dot V=\dot{\delta V}=(2C)^{-1}(I_L-I_{\rm SQ})$. In the leading order of the perturbative expansion, $I_L\sim -\bar V/R$ and we can approximately set $V\sim\bar V$ and $\tilde\varphi(t)\sim 2e\bar Vt/\hbar$ in $I_{\rm SQ}(\bar V,\tilde\varphi,\Phi)$. We obtain
\begin{equation}
\dot{\delta V}\simeq -\frac{2|\cos(\pi\frac{\Phi}{\Phi_0})|}{2C}\left[I_{ j}(\bar V)\sin\left(\frac{2e\bar Vt}{\hbar} \right)+I_{\rm int}(\bar V)\cos\left(\frac{2e\bar Vt}{\hbar}\right)\right].
\end{equation}
Since $I_j(\bar V)\gg |I_{\rm int}(\bar V)|$, we can neglect the first term in the right side of the equation in first approximation, and obtain by integration
\begin{equation}
V(t)\simeq\bar V + \frac{\hbar}{4e |\bar V| C}I_{j}(\Phi)\cos\left(\frac{2e\bar Vt}{\hbar} \right)
\label{eq:pertV}
\end{equation}
where $I_{j}(\Phi)=2I_{j}(\bar V)|\cos(\pi\Phi/\Phi_0)|$~\footnote{For $\alpha\lesssim 1$, one can substitute $|\cos(\pi\Phi/\Phi_0)|\rightarrow 0.5 \sqrt{1+\alpha^2 +2\alpha \cos(2\pi\Phi/\Phi_0)}$ in first approximation.} is the amplitude of the Josephson current of the SQUID. The validity of the perturbative solution is good when the size of the correction is much smaller then the leading term, i.e., $\hbar I_{j}(\Phi)/(4e |\bar V| C)\ll |\bar V|$. In terms of the Josephson current suppression, the previous relation requires $I_{j}(\Phi)/I_{\rm J}(\Phi=0)\ll 4eC |\bar V|^2/[\hbar I_{j}(\Phi=0)]$. The typical thermo-voltage is of order $\bar V\sim 0.1\Delta_{0,S}/e$, whereas the critical current is roughly $I_{j}(\Phi=0)\sim G_T\pi \Delta_{0,S}/e$ , giving $I_{j}(\Phi)/I_{j}(\Phi=0)\ll 0.01 C\Delta_{0,S}/(G_T\hbar)$. Interestingly, this last inequality shows that the requirement on the Josephson coupling suppression to generate an average thermoelectric signal depends on the superconducting gap but not necessarily on the geometric area of the junction, since both $C$ and $G_T$ are proportional to the area of the junction. For an aluminum based structure, characterized by $\Delta_{0,S}\sim 0.2$ meV, specific capacitance of the barrier $C/A =50$ fF/$\mu$m$^2$ and specific conductance $G_T/A =1$ mS/$\mu$m$^2$, one obtain $I_{j}(\Phi)/I_{j}(\Phi=0)\ll 0.15$ in the worst case scenario. This means that sometimes a moderate suppression of the Josephson coupling is sufficient to generate a dc thermovoltage.
\subsubsection{Flux evolution of the circuit dynamics}
\label{sec:Fluxevolution}
Now we wish to discuss the crossover between $\Phi=0$, where the phase-coherent contribution is maximum, to the case $\Phi=0.5\Phi_0$, where it is zero. We consider a few cases where $R> V_p/[2I(V_p)]$, thus the load is large enough to induce a spontaneous symmetry breaking in the absence of Josephson contributions (see colored points in Fig.~\ref{Fig2}a). We solved numerically the system of Eqs.\ref{sys:circuit}: after a transient evolution, the steady-state solution is periodic, with a period $T$ which depends both on the load and on the SQUID magnetic flux. Figure~\ref{Fig2}b displays the average value of the voltage signal in the steady-state $\bar V=1/T\int_{t_0}^{t_0+T}V(t')dt'$ as a function of the magnetic flux for the three values of the load considered. For all the curves, the mean voltage is finite and very close to the value at $\Phi=0.5\Phi_0$ for values of the flux where the critical current is strongly suppressed. These results show that even in the presence of a small phase-coherent (Josephson) contribution, the junction may still generate a breaking of EH symmetry and a net dc thermoelectric contribution. On the other hand, for large values of the critical current the mean voltage drops to zero. Note that the critical value of the flux where the $\bar V$ switch from zero to a finite value depends on various parameters, and in particular on the load resistance. For a large load, the dc thermoelectric voltage is present even in the presence of a moderate Josephson current. In order to characterize the dynamics more completely, we computed the frequency $f=1/T$ and the amplitude, defined as $\mathcal A_V=[\max_{t} V(t)-\min_t V(t)]/2$, of the voltage oscillations in the steady-state. Figure~\ref{Fig2}c displays the flux evolution of the frequency. In particular, the frequency decreases by reducing the critical current in the region where the average value of the voltage is zero, whereas it is larger and it is exactly proportional to the mean value $\bar V$ of the oscillations when $\bar V\neq 0$, due to the ac Josephson effect. The corresponding amplitude of the oscillations is shown in Fig.~\ref{Fig2}d. In the region where the $\bar V=0$, the amplitude slightly decreases upon decreasing the phase-coherent contributions, and then decrease sharply to a small value in the proximity of $\Phi=0.5\Phi_0$, where the amplitude is well described by the prefactor of the cosine term in Eq.~\eqref{eq:pertV} (see dashed lines in Fig.~\ref{Fig2}d). 
\subsection{Load dependence}
\label{sec:Load}
Here we give a more general discussion of the impact of the load $R$ on the dynamics of the junctions and hence on the thermoelectric features. We will consider two extreme values of the flux: $\Phi=0$, where the phase contributions is maximum and $\Phi=0.5\Phi_0$, where it is minimum. Figure~\ref{Fig3}a displays the frequency of the steady-state oscillations as a function of the load resistor. The corresponding amplitude of the voltage oscillations is shown in Fig.~\ref{Fig3}b.

\subsubsection{Small Josephson contribution.} Consider first the case $\Phi=0.5\Phi_0$, where the Josephson coupling is negligible. We can identify three main regions. For a small load, i.e., for $R\leq 10 \Omega$, the voltage bias falls mainly across the inductor $L$, i.e., $V(t)\sim -L\dot I_L(t)$ and the system behaves as a $LC$ oscillator of characteristic frequency $f\sim f_{\rm LC}=(2\pi\sqrt{2LC})^{-1}\sim 11.3$ GHz. The steady-state oscillations are characterized by a zero mean value of the voltage bias and a sizable amplitude which can be computed through the energy balance in the system. In particular, in the steady-state, the total energy dissipated in one cycle in the resistor (Joule heating) must be equal to the total energy provided by the superconducting junctions,
\begin{equation}
\int_{0}^{T}Ri_{L}(t')^2dt'=-\int_0^T I_{\rm SQ}(t')V(t')dt'.
\label{eq:enbalancegeneral}
\end{equation}
In the steady-state, the previous equation is generally valid irrespectively on the strength of the Josephson coupling. In the case considered here ($\Phi=0.5\Phi_0$), the SQUID current is given by the quasiparticle transport, $I_{\rm SQ}(t')=2 I_{\rm qp}(V(t'))$. In general, the power generated by the thermoelectric effects in the junction is able to self-sustain an oscillatory behaviour in the steady-state. It is relevant to note that the thermoelectric power is given both by $I_{\rm qp}$ and $I_{\rm int}$ (if present). Assuming a quasi-sinusoidal oscillatory regime in the steady-state with zero average value, i.e., $V(t)\simeq \mathcal A_V\cos(2\pi f_{\rm LC} t)$, $I_L(t)\simeq \mathcal A_I\sin(2\pi f_{\rm LC} t)$ [with $\mathcal A_I=-\mathcal A_V/(2\pi f_{\rm LC}L)$ since $V(t)\sim -L\dot I_L(t)$], the energy balance Eq.~\eqref{eq:enbalancegeneral} yields an integral-algebraic equation in $A_V$ which can be numerically solved for each value of $R$. The result of this approximation is shown in Fig.~\ref{Fig3}b with a double-dotted dashed curve (lower) and describes very accurately the amplitude of the steady-state signal computed through the numerical solution of the differential equations in Eq.~\eqref{sys:circuit}.

At intermediate loads, i.e., $10\Omega\leq R\leq 200\Omega $, the system shows a pure-dissipative behaviour and relaxes to a time independent zero-current state. We stress that in the range $150\Omega\leq R\leq 200 \Omega$, the junction also supports time-independent solutions with finite voltage [see dashed line in Fig.~\ref{Fig2}a, which represents the solution of the implicit Eq.~\eqref{eq:implicit}], and the time-evolution can be either dissipative and lead to the zero-current solution $I_L=V=0$ in the steady-state or may produce a finite dc-thermoelectric voltage, depending on the particular initial conditions. 

For large loads, $R>200\Omega$, the system oscillates around the thermoelectric solution $\bar V$ [see solid curve in Fig.~\ref{Fig2}a] with frequency $f=\bar V/\Phi_0$ and a small amplitude. The latter is better visualized in the inset of Fig.~\ref{Fig3}b, which shows a magnification of the amplitude, which for this example is of order $1\mu$V. Note that the expression is well described by the coefficient of the cosine term in Eq.~\eqref{eq:pertV}, displayed in the inset with a dashed curve. Thus, the frequency increases monotonically with the load [since the average thermoelectric voltage $\bar V$ is monotonically increasing, see Fig.~\ref{Fig2}a] and the amplitude is moderately decreasing.

\subsubsection{Large Josephson contribution.}
\begin{figure}[tp]
	\begin{centering}
		\includegraphics[width=\columnwidth]{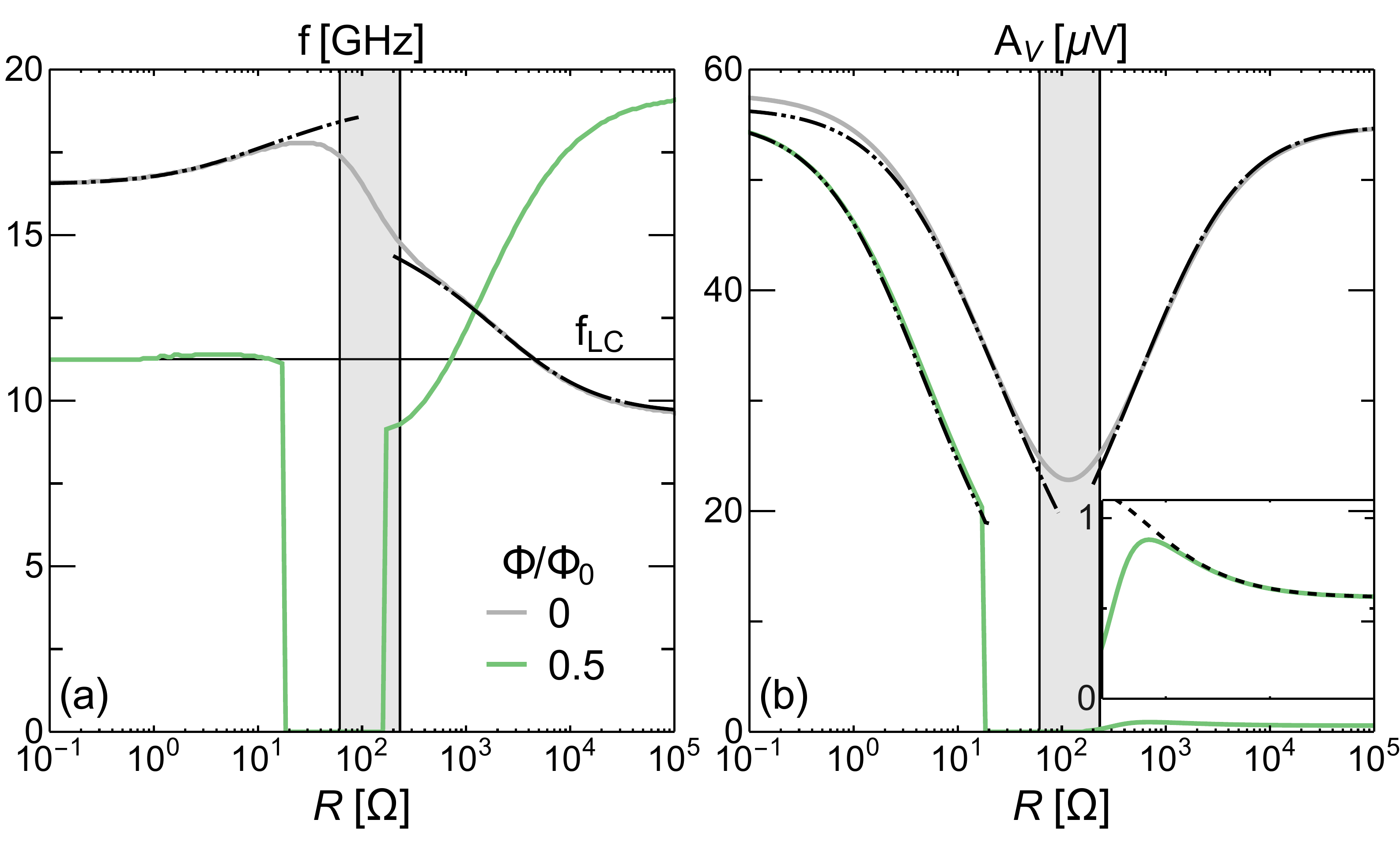}
		\caption{\textbf{(a)-(b)} Load evolution of the frequency (a) and the amplitude (b) of the steady-state oscillations for minimal and maximal phase-coherent contributions. The gray region denotes the area where the system displays a chaotic behaviour for $\Phi=0$, which may result in a relaxation to a zero-current time-independent solution. The plot also displays the approximation expressions for the frequency and the amplitude of the oscillation in the low-load limit (double-dotted dashed) and the large-load limit (dotted dashed). (see the appendix for the details of the calculations). Parameters are: $G_{T}^{-1}=1{\rm k}\Omega$, $L=1$ nH,  $C=100$ fF, $T_{S}=0.7T_{c,S}$, $T_{S'}=0.01T_{c,S}$, $T_{c,S}=1.6 $K, and $r=0.75$.}
		\label{Fig3}
	\end{centering}
\end{figure}

For a large value of the Josephson current, the evolution is qualitatively different. In particular, the frequency is monotonically increasing for low values of the load $R<20\Omega$, where it reaches a maximum $f_{\rm max}\sim 17$ GHz, and monotonically decreasing at larger values. The amplitude of the oscillations follows the inverse pattern, with a monotonically decreasing evolution for $R<100\Omega$, and a growth at larger values. It is possible to compute numerically with a good degree of approximation the load evolution of the frequency and the amplitude of the voltage oscillations (without integrating explicitly Eq.\eqref{sys:circuit}) both in the low-load limit (roughly for $R\leq 10 \Omega$, see the upper double-dotted dashed curves in Fig.\ref{Fig3}a and in Fig.\ref{Fig3}b) and in the large load limit ($R>200 \Omega$, see the upper dotted dashed curves in Fig.\ref{Fig3}a and in Fig.\ref{Fig3}b). Here we give the fundamental elements of the theoretical approach, and we leave a more detailed discussion of the modeling to the Appendix C.

In the low-load limit, the frequency of the oscillations is increased with respect to the zero Josephson coupling case. In fact, the system still behaves approximately as a LC oscillator, but with a modified inductance $L_{\rm eff}$, which is the parallel of the external inductor $L$ and the Josephson inductance~\cite{Likharevbook} $L_j=\Phi_0/[2\pi 2I_j(0)]\sim 0.55$ nH, namely $L_{\rm eff}=(L^{-1}+L_j^{-1})^{-1}\sim 0.35$ nH and characteristic frequency $f\sim 19$GHz. Note that the actual value of the frequency is slightly smaller and dependent on the load $R$. This behavior is related to the nonlinear phase dynamics of the junctions, which is associated with a frequency and amplitude dependence of the effective inductance of the circuit $L_{\rm eff}(f,\mathcal A_V)$, as shown in Appendix C. Therefore, the approximate expressions for $f,\mathcal A_V$ (upper double-dotted dashed curves in Fig.~\ref{Fig3}) are obtained by solving self-consistently for $ f=[2\pi \sqrt{2 L_{\rm eff}(f,\mathcal A_V) C}]^{-1}$ and the energy balance Eq.~\eqref{eq:enbalancegeneral} in the circuit. In the latter, one can obtain an integral equation in terms of $f,\mathcal A_V$ by assuming a quasi-sinusoidal regime, similarly to the zero Josephson coupling case. Moreover, the Cooper pairs term $I_j(V)$ plays no role in the energy balance since it is purely reactive, and only affects the effective inductance of the circuit  $L_{\rm eff}(f,\mathcal A_V)$, as discussed above.  

For a large load, the voltage drop occurs mostly in the resistor $I_L(t)\sim -V(t)/R$, and we can write the first of Eq.~\eqref{sys:circuit} as a second order nonlinear differential equation in the phase-bias
\begin{equation}
C\ddot{\tilde\varphi}+\frac{2\pi}{ \Phi_0}I_j\left[\frac{\dot{\tilde\varphi}\Phi_0}{2\pi}\right]\sin\tilde\varphi=\frac{-\pi\mathcal F (\tilde\varphi,\dot{\tilde\varphi})}{\Phi_0}
\label{eq:genpendulum}
\end{equation}
where the effective external forces are
\begin{equation}
\mathcal F (\tilde\varphi,\dot{\tilde\varphi})=\frac{\Phi_0}{2\pi R}\dot{\tilde\varphi} +2I_{\rm qp}\left[\frac{\dot{\tilde\varphi}\Phi_0}{2\pi}\right]+2I_{\rm int}\left[\frac{\dot{\tilde\varphi}\Phi_0}{2\pi}\right]\cos\tilde\varphi.
\label{eq:disspendulum}
\end{equation}
For $|V(t)|=|\Phi_0\dot{\tilde\varphi}(t)/2\pi|\ll V_p$, one can approximate $I_{j}(V)\simeq I_{j}(V=0)$ (see Fig.\ref{Fig1}b), and Eq.~\eqref{eq:genpendulum} yields a damped pendulum equation with zero-amplitude angular frequency $\omega_0^2=1/(2L_j C)$ and additional nonlinear terms which involves both damping and power generation in the presence of the thermoelectric effect. Unfortunately, in the typical situation considered here, the amplitude of the voltage oscillations is larger than $V_p$ and the approximation $I_{j}(V)\simeq I_{j}(V=0)$ is inaccurate. Hence, the system behaves as a nonlinear pendulum where the zero-amplitude frequency depends on $V\propto \dot{\tilde\varphi}$ [in the mechanical pendulum analogue, the length changes during the evolution similar to an elastic string]. The right-hand side of Eq.~\eqref{eq:genpendulum} contains both the damping and the driving force of the nonlinear pendulum. In particular, in Eq.~\eqref{eq:disspendulum} the first term in the right-hand side gives the damping associated to the Joule heating in the load. The second term gives the quasiparticle current, which is active when $|V|< V_S$ and dissipative otherwise, where $V_S\neq 0$ is the Seebeck voltage, [we recall that at the Seebeck voltage the quasiparticle current is zero $I_{\rm qp}(V_S)= 0$]. A similar behavior, active at low voltage bias and dissipative at higher voltage bias, applies also to the interference term.

In order to evaluate $f,\mathcal A_V$ as a function of the load $R$, one has to solve self-consistently the energy-balance in the steady-state Eq.~\eqref{eq:enbalancegeneral} [which involves the terms in Eq.~\eqref{eq:disspendulum}] and the relation between the frequency and the amplitude in the nonlinear pendulum~\cite{Mickens1996},
\begin{equation}
\omega=2\pi f=\frac{\pi}{2}\omega_0\frac{1}{K[\sin\mathcal A_\varphi/2]}.
\label{eq:freqnonlinearpendulum}
\end{equation}
Above, $K[k]$ is the complete elliptic integral of the 1st kind, $\mathcal A_{\tilde\varphi}=[\max_{t} \tilde\varphi(t)-\min_t \tilde\varphi(t)]/2$ is the amplitude of the phase oscillations, and we replaced $I_{j}(0)\to \mathcal A_V^{-1}\int_0^{\mathcal A_V} I_{j}(V')dV'$ in $\omega_0=1/\sqrt{2L_jC}$. The theoretical modeling exploits an highly accuracy approximate solution of the nonlinear pendulum equation~\cite{BELENDEZ} which includes the effect of higher harmonics (see Appendix C for a detailed discussion), and perfectly describes the motion of the system (see the upper dotted-dashed curves in Fig.~\ref{Fig3}). The load evolution of $f,\mathcal A_V$ can be qualitatively understood as follows. By increasing $R$, the dissipation in the circuit for a given voltage bias $V(t)$ is reduced, since $RI_L(t)^2\sim V^2(t)/R$, producing an increase in the amplitude of the oscillations. As a consequence, the frequency of the oscillations decreases, since in the nonlinear pendulum the frequency is monotonically decreasing with the amplitude of the oscillations [see Eq.~\eqref{eq:freqnonlinearpendulum}].

Finally, we note that the behaviour of the junction is chaotic at intermediate values of $R$ (see filled regions). In particular, the system may relax to a zero-current time independent solution, depending on the initial conditions. This can be understood by inquiring the eigenvalues of the linearized equations which describe the dynamics of the system close to the stationary solutions (see Appendix B).
\section{Conclusions And Discussion} 
In summary, we have discussed the dynamics of thermally biased Josephson junctions, in the presence of the nonlinear thermoelectric effect recently predicted in tunnel junctions between two different BCS superconductors. We investigated a system where the size of the Josephson coupling can be externally tuned, by modulating the flux inside a SQUID. The system displays a rich phenomenology, when inserted in a generic electric circuit, such as a RL circuit. Depending on the load, we focused on two relevant different regimes. In the presence of a large load, the system generates a finite dc-thermoelectric voltage when the Josephson coupling is strongly suppressed but still finite, due to the spontaneous breaking of EH symmetry. In addiction, the system outputs an ac signal with frequency exactly proportional to the thermoelectric voltage, due to the ac Josephson effect. As a consequence, both the thermoelectric voltage and the ac signal can be ultimately controlled by changing the size of the load. When the Josephson coupling is stronger, the system generates a pure ac-thermoelectric signal. When the load connected to the system is small, the systems generates an ac signal, independently on the strength of the Josephson coupling. Interestingly, the modulation of the Josephson current induces a control of the effective inductance of the circuit, and hence of the frequency of the thermoelectric signal. The operating ranges depend on the inductance connected to the circuit and are in the GHz regime for a standard aluminum based structures. We may envision different applications for this system, taking advantage of the different regimes. Firstly, we note that when the system generates a dc thermoelectric signal, one has an autonomous system that convert a temperature gradient in a dc voltage signal which is perfectly tuned (by the Josephson relation) with the frequency of the ac component. This may find some value when one need to have a controlled generator that need to be galvanically disconnected from external circuits. Another application may involve the detection of radiation: by tuning the system very close to the transition point where the mean thermoelectric voltage switches from $\bar V=0$ to $\bar V\gtrsim V_p$, the system is highly sensitive to small parameters variation, such as the load or the temperature difference. Therefore, events such as photon absorption may trigger the spontaneous breaking of EH symmetry. Finally, one may envision an application as an high frequency oscillator controlled by the flux and feed with a thermal gradient only. We believe that the discussed system presents novel properties and functionalities that can be relevant in the field of superconducting quantum technologies.

\begin{acknowledgments}
We acknowledge the EU's Horizon 2020 research and innovation programme under grant agreement No. 800923 (SUPERTED) for
partial financial support. A.B. acknowledges the CNR-CONICET
cooperation program "Energy conversion in quantum nanoscale hybrid devices", the SNS-WIS joint lab QUANTRA funded by the Italian Ministry of Foreign Affairs and International Cooperation, and the Royal Society through the International Exchanges between the UK and Italy (Grants No. IES R3 170054 and IEC R2 192166).
\end{acknowledgments}

\appendix

\section{Microscopic expressions of the tunneling current}
The tunneling expressions of Eq.~\eqref{eq:IVphi_characteristic} are
\begin{widetext}
\begin{align}
I_{\rm qp}(V)&=\frac{G_{T}}{e}\int_{-\infty}^{+\infty}dE N_{S}(E) N_{S'}(E-eV)[f_{S'}(E-eV)-f_{S}(E)]
\label{eq:Iqp}\\
I_{j}(V)&=-\frac{ G_{T}}{2e}\int_{-\infty}^{+\infty}dE\{\Re [ F_{S}(E)]\Im[ F_{S'}( E-eV)]f_{S}(E)+ \Re [ F_{S'}(E-eV)]\Im[ F_{S}(E)]f_{\rm S'}( E-eV)]\}\label{eq:Ij}\\
I_{\rm int}(V)&=\frac{ G_{T}}{e}\int_{-\infty}^{+\infty}dE \Re[ F_{S}(E)] \Re [ F_{S'}(E-eV)][f_{\rm S'}( E-eV)-f_{S}(E)]\label{eq:Iint}
\end{align} 
\end{widetext}
where $\Re[\dots] $ and $\Im[\dots]$ denote the real and the imaginary parts, respectively. In the BCS model, the quasiparticle density of states read $N_{i}(E)=|\Re[(E+j\Gamma_i)/\sqrt{(E+j\Gamma_i)^2-\Delta^2_i}]|$ and $ F_{i}(E)={\rm sgn}(E)\Delta_{i}/\sqrt{(E+ j\Gamma_i)^2-\Delta^2_i}$ are the anomalous Green's functions (here $i={S,S'}$ and $j$ is the imaginary unit). We assumed the electrodes in the quasi-equilibrium regime, hence the quasiparticle distributions are the Fermi functions $f_i(E)=[\exp(E/k_B T_i)+1]^{-1}$, where $k_B$ is the Boltzmann constant. The phenomenological parameters $\Gamma_i$ (typically called Dynes parameters) give a phenomenological representation of the finite quasiparticle lifetime~\cite{Dynes1978,Dynes1984} or the influence
of the electromagnetic environment of a tunnel junction~\cite{PekolaDynes}. In all the calculations, we set $\Gamma_i=10^{-4}\Delta_{0,i}$. Equation~\eqref{eq:IVphi_characteristic} (with the expressions Eqs.~\eqref{eq:Iqp},\eqref{eq:Ij}, and \eqref{eq:Iint}) are derived in the tunneling limit for a constant voltage bias $V$~\cite{barone1982physics,Harris1974}. In the presence of a time dependent voltage, the expression of Eq.~\eqref{eq:IVphi_characteristic} does not hold generally anymore, and must be generalized to include time-delayed effects also~\cite{barone1982physics,HarrisPRB11}. However, in this work we consider the adiabatic regime~\cite{barone1982physics,Likharevbook}, where we can still use the expression of Eq.~\eqref{eq:IVphi_characteristic}, replacing $V\to V(t)$. The adiabatic approximation holds when the voltage signal is small $eV(t)\ll \Delta_{\rm 0,S}$, or the time variations of $V(t)$ are small compared to the gap frequency $\sim(1+r)\Delta_{0,S}/\hbar$~\cite{barone1982physics,Likharevbook}. We consider realistic values of the circuit parameters where both these conditions are reasonably fulfilled. We expect that the main predictions are not crucially affected even beyond the adiabatic approximations. As discussed in Refs.~\onlinecite{MarchegianiPRL,MarchegianiPRB}, the quasiparticle current shows a thermoelectric behaviour, i.e., $I_{\rm qp}(V)V<0$ for $T_S>T_S'$, provided $\Delta_{\rm S}(T_S)>\Delta_{S'}(T_S')$. In the limit $T_S'\to 0$ and for small values of the bias $V\rightarrow 0$, the currents is approximately linear $I_{\rm int}\sim G_{\rm 0,qp}V$, where the zero-bias differential conductance reads~\cite{MarchegianiPRL,MarchegianiPRB}
\begin{equation}
G_{\rm 0,qp}=
-2G_T\Delta_{0,S'}^2\int_{\Delta_{S}(T_S)}^{\infty}dE \frac{N_{S}(E)f_{\rm S}(E)}{(E^2-\Delta_{0,S'}^2)^{3/2}}.
\label{eq:G0qp}
\end{equation} 
A similar expression can be derived for the quasiparticle interference term in the same limit, where $I_{\rm int}\sim G_{\rm 0,int}V$, with 
\begin{equation}
G_{\rm 0,int}=
-2G_T\Delta_{0,S'}\Delta_{S}(T_S)\int_{\Delta_{S}(T_S)}^{\infty}dE \frac{N_{\rm S}(E)f_{\rm S}(E)}{(E^2-\Delta_{0,S'}^2)^{3/2}}.
\label{eq:G0int}
\end{equation} 
Note that the ratio of the two quantities is given by the expression quoted in the main text,
\begin{equation}
\frac{G_{\rm 0,int}}{G_{\rm 0,qp}}= \frac{\Delta_{S}(T_S)}{\Delta_{0,S'}}\geq 1
\end{equation}
where the inequality holds in the thermoelectric regime, where $\Delta_{S}(T_S)\geq\Delta_{0,S'}$.
\section{Linearization and stability analysis}
\begin{figure}[h]
	\begin{centering}
		\includegraphics[width=\columnwidth]{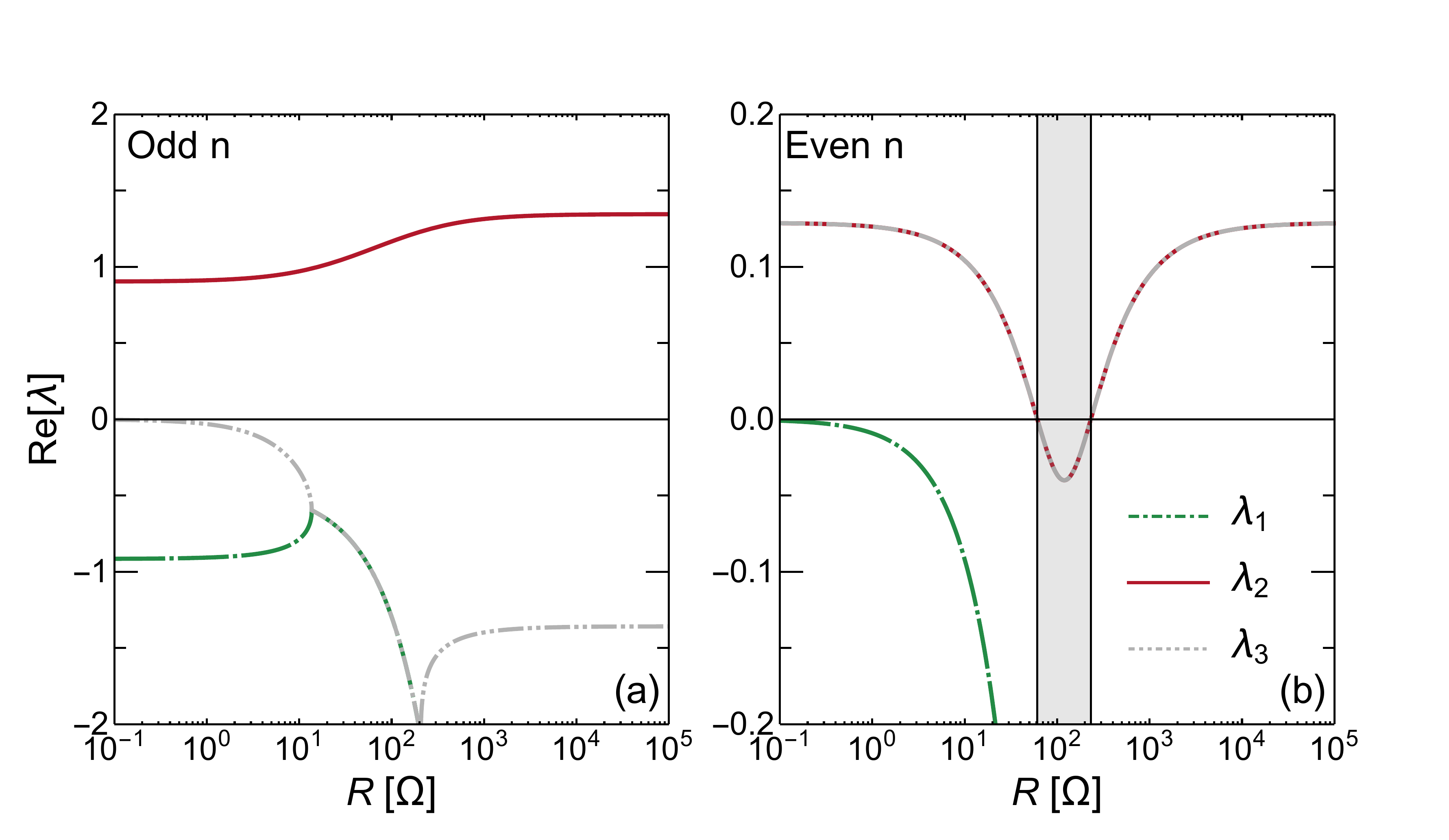}
		\caption{Evolution of the real part of the eigenvalues of the matrix $\mathbb M$ in Eq.~\eqref{eq:matrix} for odd values \textbf{(a)} and for even values \textbf{(b)} of $n$ (for $\tilde\Phi=0$). The filled region in panel (b) denotes the interval where all the eigenvalues have negative real part, and so the stationary point solution is stable.
		}
		\label{FigB1}
	\end{centering}
\end{figure}

In order to describe the different regimes of the dynamical system, it is convenient to work in scaled units, namely we consider $i_{i}=eI_{i}/G_{\rm T}\Delta_{0,S}$ (with the subscript $i=\{L,{\rm SQ}\}$), $v=eV/\Delta_{0,S}$, $\tilde\Phi=\pi\Phi/\Phi_0$ and $\tau=t/\sqrt{2LC}$. The frequency of the oscillations are obtained by multiplying the scaled frequency $\tilde\omega$ by $f_{\rm LC}=(2\pi\sqrt{2LC})^{-1}\sim 11.3$ GHz for our parameters choice. The system of Eq.~\eqref{sys:circuit} in scaled units reads 
\begin{equation}
\begin{cases}
\dot v=\epsilon \{i_{L}-2i_{\rm qp}(v)-2|\cos\tilde\Phi|[i_{ j}(v)\sin\left(\tilde\varphi\right)+i_{\rm int}(v)\cos\left(\tilde\varphi\right)]\} \\
\epsilon\dot i_{L}=-\xi i_{L}-v\\
\dot {\tilde\varphi}=\kappa v.
\end{cases}
\label{sys:scaledcircuit}
\end{equation}
Note that the dynamics of the system depends on three dimensionless parameters: $\epsilon,\kappa,\xi$. More precisely,  $\kappa=2\Delta_{0,S}\sqrt{2LC}/\hbar$ is the ratio between the gap frequency $2\Delta_{0,S}/\hbar$ and the angular frequency of the LC oscillations $2\pi f_{LC}$. As discussed in Appendix A, the validity of Eq.~\eqref{eq:IVphi_characteristic} [with the expressions Eqs. ~\eqref{eq:Iqp},~\eqref{eq:Ij}, and \eqref{eq:Iint}] is restricted to the adiabatic regime, where the time variations are much smaller than the gap frequency, i.e., $\kappa\gg 1$ [in our calculation, we set $\kappa\sim 10.5$]. 
The other parameters are: $\xi=G_{T}R=R/R_T$, which is the ratio between the load and the normal state resistance $R_T$ and $\epsilon=G_{T}\sqrt{L/(2C)}$, which is proportional to the strength of the thermoelectric effect and thus also characterizes the coupling to the nonlinear terms of the system of equations. It is convenient to have a small value of $\epsilon$, to avoid strong non-linearities in the dynamics [in the calculations, we set $\epsilon\sim 0.07$]. It is worth noting that the values of $\kappa$ and $\epsilon$ adopted are obtained by considering realistic values for typical Josephson junctions realized through standard nanofabrication techniques.  The stationary and time independent solutions are obtained by setting $\dot v=\dot{\tilde\varphi}=\dot i_L=0$ and read $ v= i_{L}=0,\tilde\varphi=n\pi$ (with $n\in \mathbb{Z}$). The stability analysis can be inquired with a standard linearization procedure, which leads to the matrix equation
\begin{widetext}
\begin{equation}
\begin{pmatrix}
\dot v   \\   
\dot i_{L}  \\
\dot{\tilde\varphi}  
\end{pmatrix}
=
\begin{pmatrix}
-2\epsilon[g_{\rm qp}+ (-1)^{n}g_{\rm int}|\cos\tilde\Phi|]   & \epsilon & -(-1)^n 2\epsilon i_{j,0}|\cos\tilde\Phi|\\   
-\epsilon^{-1}   &  -\xi/\epsilon & 0 \\
\kappa   &  0 & 0   
\end{pmatrix}
\begin{pmatrix}
v   \\   
i_{L}  \\
\tilde\varphi  
\end{pmatrix}=\mathbb M \begin{pmatrix}
v   \\   
i_{L}  \\
\tilde\varphi  
\end{pmatrix}
\label{eq:matrix}
\end{equation}
\end{widetext}
where $g_{i}=G_{i}(V=0)/G_T$ (with $i=\{{\rm qp,int}\}$) is the scaled zero-bias differential conductance and $i_{j,0}=i_j(v=0)$. In particular, a necessary condition for the stability of the stationary and time independent solutions is that the real parts of all the eigenvalues of the matrix in Eq.~\eqref{eq:matrix} must be negative~\cite{strogatz}. The eigenvalues $\lambda$ can be obtained by solving the characteristic equation, obtained by setting  $\det{(\mathbb M-\lambda \mathbb I)}=0$,
\begin{widetext}
\begin{equation}
\lambda^3+[2\epsilon(g_{\rm qp}+(-1)^{n}g_{\rm int}|\cos\tilde\Phi|)+\xi/\epsilon]\lambda^2
+[1+2g_{\rm qp}\xi+(-1)^{n}|\cos\tilde\Phi|(g_{\rm int}\xi+i_{j,0}\kappa\epsilon) ]\lambda+(-1)^{n}2\kappa\xi i_{ j,0}|\cos\tilde\Phi|=0.
\end{equation}
\end{widetext}
where $\mathbb I$ is the $3\times3$ identity matrix. The explicit expressions of the eigenvalues in terms of the various parameters of the system are obtained by using the cubic formula (not shown here). Figure~\ref{FigB1} displays the load evolution of the real part of the eigenvalues $\lambda_{1,2,3}$ as a function of the load resistor for the set of parameters used in the main text, both for odd values [Fig.~\ref{FigB1}a] and even values [Fig.~\ref{FigB1}b] of $n$. Note that, for odd values of $n$ (Fig.~\ref{FigB1}a) the real part of $\lambda_{2}$ (solid red) is positive irrespectively of $R$. As a consequence, the stationary time independent solutions $ v= i_{L}=0,\tilde\varphi=n\pi$ (with odd $n$) are always unstable. The situation is different for even values of $n$ [Fig.~\ref{FigB1}b]. In particular, the plot shows that for $60 \Omega\leq R\leq 230\Omega$ (filled region) all the real parts of the eigenvalues are negative [the real parts of $\lambda_{2}$ and $\lambda_{3}$ coincide since $\lambda_{2}=\lambda_{3}^*$ in this case]. As a consequence, in this region the stationary time dependent solution characterized by $ v= i_{L}=0,\tilde\varphi=n\pi$ (with even $n$) is stable, and the time dependent evolution of the system depends on the initial conditions, as we verified numerically by solving Eq.~\eqref{sys:scaledcircuit} for different values of $i_L(\tau=0),v(\tau=0),\tilde\varphi(\tau=0)$. More precisely, if the system is slightly perturbed around the stationary solution, the evolution relaxes to it. For larger perturbations, the systems approach a limit cycle characterized by a periodic oscillation whose amplitude and frequency are shown in Fig.~\ref{Fig3} of the main text.
\section{Strong Josephson contribution}
In the presence of a strong Josephson current, the electron-hole symmetry breaking is only obtained in the time-dependent domain, and both the mean value of the current $\bar i$ and the mean voltage $\bar v$ are equal to 0. We focus on two different regimes, related to the value of the load. For simplicity, we consider the case of zero-flux, but the results can be extended to $\Phi\neq 0$.
\subsection{Small Load}
In the presence of a small load, we look for a perturbative solution for the current in the circuit 
\begin{equation}
i_L(\tau)=i_L^{(0)}(\tau)+\xi i_L^{(1)}(\tau)+\dots
\label{eq:curr-lowR}
\end{equation}
where $i_L^{(0)}(\tau)$ is the solution in the absence of the load $\xi\to0$.
Inserting in the second of Eqs.~\eqref{sys:scaledcircuit}, we can extract by perturbative analysis in $\xi$
\begin{align}
&\epsilon \dot i_{ L}^{(0)}= -v\nonumber\\
&\epsilon \dot i_{ L}^{(1)}=-i_L^{(0)}.
\end{align}
Upon insertion in the first equation of Eq.~\eqref{sys:scaledcircuit}, we obtain
\begin{equation}
\ddot i_{L}^{(0)}+i_{L}^{(0)}+\xi i_L^{(1)}=2i_{\rm qp}(v)+2i_{j}(v)\sin\left(\tilde\varphi\right)+2i_{\rm int}(v)\cos\left(\tilde\varphi\right)
\end{equation}
We are interested in the steady-state oscillatory evolution of the system, characterized by an unknown angular frequency $\tilde\omega=2\pi/\tilde T$ (here $\tilde T$ is the scaled period). By assuming a quasi-sinusoidal oscillation in the voltage, neglecting higher harmonics of the oscillations, one gets
\begin{align}
v(t)&=\tilde{\mathcal A_{V}}\sin(\tilde\omega t)\nonumber\\
\dot{\tilde\varphi}=\kappa v\rightarrow \tilde\varphi(t)=-\frac{\kappa\tilde{\mathcal A_{V}}}{\tilde\omega}\cos(\tilde\omega t)=
\tilde\varphi(t)&=-\tilde{\mathcal A}_{\tilde\varphi}\cos(\tilde\omega t)
\nonumber\\
\epsilon \dot i_L^{(0)}(t)=-v(t)\rightarrow i_{ L}^{(0)}(t)&=\frac{\tilde{\mathcal A_{V}}}{\epsilon\tilde\omega}\cos(\tilde\omega t)\nonumber\\
\epsilon \dot i_L^{(0)}(t)=-\xi i_L^{(0)}(t)\rightarrow i_{ L}^{(1)}(t)&=-\frac{\tilde{\mathcal A_{V}}}{\epsilon^2\tilde\omega^2}\sin(\tilde\omega t)\nonumber\\
\end{align}
where we defined $\tilde{\mathcal A}_{\tilde\varphi}=\kappa\tilde{\mathcal A_{V}}/\tilde\omega$.
In order to compute $\tilde{\mathcal A_{V}},\tilde\omega$, we insert these expression in the current conservation equation Eq.~\eqref{eq:curr-lowR}, and obtain
\begin{widetext}
\begin{equation}
\frac{\tilde{\mathcal A_{V}}}{\epsilon\tilde\omega}(-\tilde\omega^2+1)\cos\psi-\frac{\xi\tilde{\mathcal A_{V}}}{\epsilon^2\tilde\omega^2}\sin\psi=2 \left[i_{\rm qp}(\tilde{\mathcal A_{V}}\sin\psi)-i_{\rm j}(\tilde{\mathcal A_{V}}\sin\psi)\sin\left(\tilde{\mathcal A}_{\tilde\varphi}\cos\psi\right)+i_{\rm int}(\tilde{\mathcal A_{V}}\sin \psi)\cos\left(\tilde{\mathcal A}_{\tilde\varphi}\cos\psi\right)\right].
\end{equation}
\end{widetext}
where we defined $\psi=\tilde\omega t$. We obtain two coupled equations through multiplication by either $\cos\psi$ or $\sin\psi$ and integrating over a period. 
We get
\begin{align}
\label{eq:lowR-energybalance}
&\frac{\xi\pi\tilde{\mathcal A_{V}}}{2\epsilon^2\tilde\omega^2}+\int_0^{2\pi}[i_{\rm qp}(\tilde{\mathcal A_{V}}\sin\psi)+i_{\rm int}(\tilde{\mathcal A_{V}}\sin\psi)\cos\left(\tilde{\mathcal A}_{\tilde\varphi} \cos\psi\right)]\sin\psi d\psi=0\\
\label{eq:lowR-freqbalance}
&\tilde\omega^2=1+\frac{2\epsilon\tilde\omega}{\tilde{\mathcal A_{V}}\pi}\int_0^{2\pi} i_{j}(\tilde{\mathcal A_{V}}\sin\psi)\sin\left(\tilde{\mathcal A}_{\tilde\varphi} \cos\psi\right)\cos\psi d\psi
\end{align}
where we have divided the either active/dissipative components of the current (related to $i_{\rm qp}$ and $i_{\rm int}$, in phase with the voltage bias) by the reactive component $i_j$ (shifted by $\pi/2$ with respect to the voltage bias), exploiting the different symmetries in $\tilde\varphi,v$ of the three contributions.

Equation~\eqref{eq:lowR-energybalance} is related to the energy balance in the system, since at the steady-state the energy dissipated in the load during a period must be equal to the total energy produced in the junction for each cycle. In fact, it can be rewritten in general as 
\begin{equation}
\int_0^{\bar T} \xi i_L^2(\tau)d\tau=\int_0^{\bar T} [-i_{\rm SQ}(\tau) v(\tau)]d\tau
\label{eq:energybalance}
\end{equation}
which is exactly Eq.~\eqref{eq:enbalancegeneral} in scaled units. Equation~\eqref{eq:lowR-freqbalance} gives the relation between the frequency and the amplitude of the oscillation. The Josephson current affects the effective inductance of the circuit, and produces an increased frequency of the oscillatory behaviour with respect to the case where $i_j\sim 0$. The second term in the square brackets in Eq.~\eqref{eq:lowR-freqbalance} can be interpreted as the frequency dependent correction of the circuit inductance due to the Josephson term (in units of $1/L$). In fact, for small values of the phase-oscillations $\tilde{\mathcal A_{V}},\kappa\tilde{\mathcal A_{V}}/\tilde\omega\ll 1$ (which is never properly met in our case), the integral gives a frequency independent result:
\begin{equation}
\frac{2\epsilon\tilde\omega}{\tilde{\mathcal A_{V}}\pi}\int_0^{2\pi} i_{j}(\tilde{\mathcal A_{V}}\sin\psi)\sin\left(\tilde{\mathcal A}_{\tilde\varphi}\cos\psi\right)\cos\psi d\psi\simeq 2\kappa\epsilon i_{j,0}=\frac{L}{L_j}
\end{equation}
where $L_j=\Phi_0/[2\pi 2I_j(0)]$ is the Josephson inductance, and we used $\tilde{\mathcal A}_{\tilde\varphi}=\kappa\tilde{\mathcal A_{V}}/\tilde\omega$.
Finally, the amplitude and the frequency of the oscillation are obtained by solving self-consistently Eq.~\eqref{eq:lowR-energybalance} and Eq.~\eqref{eq:lowR-freqbalance}.
\subsection{Large Load}
In the presence of a large load, we can neglect the voltage drop across the inductor and write $i_{\rm L}(t)\sim -v(t)/\xi$. Upon substitution in the current conservation equation, we can write down a pendulum-like equation with self-forcing and dissipation [Eq.~\eqref{eq:genpendulum} in scaled units]
\begin{equation}
\frac{\ddot{\tilde\varphi}}{2\kappa\epsilon}+ i_j(\dot{\tilde\varphi}/\kappa)\sin\tilde\varphi=-\left[\frac{\dot{\tilde\varphi}}{2\kappa\xi}+i_{\rm qp}(\dot{\tilde\varphi}/\kappa)+i_{\rm int}(\dot{\tilde\varphi}/\kappa)\cos\tilde\varphi\right].
\end{equation}
As discussed in the main text, the mechanical analogue of this equation is a pendulum where the pendulum length depends on the phase derivative $\dot{\tilde\varphi}$ and it is subjected to driving and dissipative forces (right-hand side of the equation). Since the sine term changes with time during the evolution, we replace $i_j(v)$ with a value averaged over the dynamics, $i_{j}(v)\to \bar i_j=\tilde{\mathcal A_{V}}^{-1}\int_0^{\tilde{\mathcal A_{V}}} i_{j}(v)dv$. 
We approximate the frequency by using its relation to the amplitude of the phase oscillations $\tilde{\mathcal A}_{\tilde\varphi}$ by the well known expression for the standard nonlinear pendulum [Eq.~\eqref{eq:freqnonlinearpendulum} in scaled units]
\begin{equation}
\tilde\omega=\frac{2\pi}{\bar T}=\frac{\pi}{2}\tilde\omega_0\frac{1}{K[\sin\tilde{\mathcal A}_{\tilde\varphi}/2]}
\label{eq:highR-freq}
\end{equation}
where $\tilde\omega_0=\sqrt{2\kappa\epsilon\bar i_j}$ and $K[k]$ is the complete elliptic integral of the 1st kind. We verified numerically that the frequency of the steady-state oscillations is well described by this expression also for our case, upon inserting the values of $\tilde{\mathcal A}_{\tilde\varphi}$ obtained from the numerical computation. The amplitude is still related to the energy-balance in the circuit in each cycle~\eqref{eq:energybalance}. In order to properly describe the oscillations even for large values of $\tilde{\mathcal A}_{\tilde\varphi}$, we exploit the high precision approximate solution of the pendulum equation with initial amplitude $\tilde{\mathcal A}_{\tilde\varphi}$~\cite{BELENDEZ}
\begin{equation}
\tilde\varphi(t)=\frac{(c_1-c_2\tilde{\mathcal A}_{\tilde\varphi}^2)\tilde{\mathcal A}_{\tilde\varphi}\cos(\tilde\omega t)}{c_1-(20+c_2)\tilde{\mathcal A}_{\tilde\varphi}^2+20\tilde{\mathcal A}_{\tilde\varphi}^2\cos^2(\tilde\omega t)}
\label{eq:Belendez}
\end{equation}
where $c_1=960$ and $c_2=49$. The energy balance equation becomes:
\begin{equation}
\int_0^{2\pi}\left[\frac{\dot{\tilde\varphi}}{2\kappa\xi}+i_{\rm qp}(\dot{\tilde\varphi}/\kappa)+i_{\rm int}(\dot{\tilde\varphi}/\kappa)\cos\tilde{\varphi}\right]\dot {\tilde\varphi} d\psi=0
\label{eq:highR-energybalance}
\end{equation}
with $\psi=\tilde\omega t$.
With the use of Eq.~\eqref{eq:Belendez} in Eqs.~\eqref{eq:highR-freq} and ~\eqref{eq:highR-energybalance}, $\tilde{\mathcal A}_{\tilde\varphi},\tilde\omega$ are obtained by solving self-consistently Eq.~\eqref{eq:highR-energybalance} and Eq.~\eqref{eq:highR-freq}. The approximation for the amplitude of the voltage oscillations, displayed in Fig.~\ref{Fig3} of the main text, are obtained from $\tilde{\mathcal A}_{\tilde\varphi}$ using the expression 
\begin{equation}
  \tilde{\mathcal A}_{V}=\frac{\tilde{\mathcal A}_{\tilde\varphi}\tilde\omega}{\kappa}\left[1+\frac{20\tilde{\mathcal A}_{\tilde\varphi}^2}{c_1-(c_2+20\tilde{\mathcal A}_{\tilde\varphi}^2)} \right],
\end{equation}
which is derived by differentiating Eq.~\eqref{eq:Belendez} and using $v=\dot{\tilde\varphi}/\kappa$. Note that, in the sinusoidal limit, the second term in the square brackets is much smaller than one and the relation between the phase and the voltage oscillations reduces to  $\tilde{\mathcal A}_{V}\simeq\tilde{\mathcal A}_{\tilde\varphi}\tilde\omega/\kappa$, as in the analysis of the previous section. In the main text, we discussed how these approximate expressions compare with the solutions obtained through the direct numerical integration of Eq.~\eqref{sys:circuit}.

%

\end{document}